\documentclass[11pt]{article}

\usepackage{amssymb,amsthm,amscd,array}
\usepackage{graphics}
\usepackage[final]{epsfig}	

\let\ssection=\section
\renewcommand{\section}{\setcounter{equation}{0}\ssection}

\newcommand{\bbR}{\mathbb{R}}

\newcommand{\bbT}{\mathbb{T}}
\newcommand{\bbC}{\mathbb{C}}

\newcommand{\Vect}{\mathrm{Vect}}
\newcommand{\Diff}{\mathrm{Diff}}
\newcommand{\Div}{\mathrm{div}}
\newcommand{\End}{\mathrm{End}}
\newcommand{\Hom}{\mathrm{Hom}}
\newcommand{\Inv}{\mathrm{Inv}}
\newcommand{\SDiff}{\mathrm{SDiff}}
\newcommand{\SO}{\mathrm{SO}}
\newcommand{\Ad}{\mathrm{Ad}}
\newcommand{\ad}{\mathrm{ad}}

\newcommand{\cA}{{\mathcal{A}}}

\newcommand{\cF}{{\mathcal{F}}}

\newcommand{\cL}{{\mathcal{L}}}

\newcommand{\fg}{\mathfrak{g}}
\newcommand{\fG}{\mathfrak{G}}

\newcommand{\fa}{\mathfrak{a}}

\chardef\s=110
\chardef\g=103

\begin{document}

\newtheorem{thm}{Theorem}[section]
\newtheorem{lem}[thm]{Lemma}
\newtheorem{cor}[thm]{Corollary}
\newtheorem{prop}[thm]{Proposition}
\newtheorem{rmk}[thm]{Remark}
\newtheorem{exe}[thm]{Example}
\newtheorem{defi}[thm]{Definition}

\def\a{\alpha}
\def\b{\beta}
\def\d{\delta}
\def\g{\gamma}
\def\om{\omega}
\def\r{\rho}
\def\s{\sigma}
\def\vfi{\varphi}
\def\vr{\varrho}
\def\vp{\varphi}
\def\l{\lambda}
\def\m{\mu}

\title{Looped cotangent Virasoro algebra and non-linear
integrable systems in dimension $2+1$}

\author{V. Ovsienko \and C. Roger}

\date{}

\maketitle

{\abstract{
We consider a Lie algebra generalizing the Virasoro algebra
to the case of two space variables.  
We study its coadjoint representation and calculate the corresponding Euler equations.
In particular, we obtain a bi-Hamiltonian system that leads to an integrable non-linear
partial differential equation.
This equation is an analogue of the
Kadomtsev--Petviashvili (of type B) equation.
}}

\bigskip


\bigskip


\thispagestyle{empty}

\section{Introduction and motivations}

This work was initiated by the the question asked to one of the authors by T. Ratiu: whether
the Kadomtsev--Petviashvili equation can be realized as an Euler equation on
some infinite-dimensional Lie group.
In order to make the above question understandable, let us start with some basic
definitions.

\subsection{The KP and BKP equations as two generalizations of KdV}

The famous Korteweg--de Vries (in short KdV) equation 
\begin{equation}
\label{KdV}
u_t=3\,u_xu+c\,u_{xxx},
\end{equation}
where $u(t,x)$ is a smooth (complex valued) function,
is the most classic example of integrable infinite-dimensional Hamiltonian system.

The Kadomtsev--Petviashvili (KP) is a ``two space variables'' generalization of KdV.
A function $u(t,x,y)$ on $\bbR^3$ of time $t$ and two space variables $x,y$ satisfies KP if
one has
\begin{equation}
\label{KP}
u_t=3\,u_xu+c_1\,u_{xxx}+c_2\,\partial_x^{-1}u_{yy},
\end{equation}
where, as usual, the partial derivatives are denoted by the corresponding variables as lower
indices, $c_1,c_2\in\bbC$ are arbitrary constants and $\partial_x^{-1}$ denotes the
indefinite integral. Of course, there is some ambiguity in this definition, so that one can
prefer to use the following form:
$$
u_{tx}=3\,u_{xx}u+3\,u_x^2+c_1\,u_{xxxx}+c_2\,u_{yy}.
$$
The constants $c_1,c_2$ are called central charges for the reasons that will be clarified
below.
Let us notice that, if $u$ does not depend explicitly on $y$ then (\ref{KP}) reduces to the
KdV with $c=c_1$.

Another example is the more recent version of the KP equation, namely so-called KP
of type B (BKP) (see \cite{KM,BK}). 
The dispersionless BKP is of the form
\begin{equation}
\label{BKP}
u_t=\a{}\,u_xu^2+
\b\left(
u_yu+u_x\partial_x^{-1}u_y
\right)
+c\,\partial_x^{-1}u_{yy},
\end{equation}
where $\a,\b$ and $c$ are some constants.

The equations (\ref{KdV})-(\ref{BKP}) are infinite-dimensional integrable (in a weak
algebraic sense, cf. \cite{RS}) systems. They correspond to infinite hierarchies of
conservation laws and an infinite series of commuting evolution equations on the space of
functions. They are interesting both for mathematics and theoretical physics.

\begin{rmk}
{\rm
The classic KP equation (\ref{KP}) should not be confused with the KP hierarchy that became
more popular than the original KP equation itself. The KP hierarchy is a family of integrable
P.D.E. obtained by an inductive algebraic construction; the KdV equation appears as the first
term in this family while the ``classic'' KP corresponds to the third term (see
\cite{RS} and \cite{AC}). For more details on the KP and BKP hierarchies see,
e.g., \cite{Hir}.
}
\end{rmk}

\subsection{Euler equations}

The notion of \textit{Euler equation} means in our context some group-theoretic
generalizations of the classic Euler equation for the rigid solid motion. 

Let $G$ be a Lie
group, $\fg$ the corresponding Lie algebra and $\fg^*$ the dual of $\fg$ equipped with the
canonical linear Poisson structure (or the Kirillov-Kostant-Souriau bracket). Consider a
linear map $I:\fg\to\fg^*$ called the
\textit{inertia} operator. 

\begin{defi}
{\rm
The Euler equation is the Hamiltonian vector field on $\fg^*$
with the quadratic Hamiltonian function 
\begin{equation}
\label{EulHam}
H(m)=\frac{1}{2}\,
\langle
I^{-1}(m),m
\rangle
\end{equation}
for $m\in\fg^*$ and $\langle\,,\,\rangle$ being the natural pairing between $\fg$ and
$\fg^*$.
}
\end{defi}
The well-known formula for this vector field is
\begin{equation}
\label{EulEq}
m_t
=
\left\{
H,m
\right\}
=
-{\ad}^*_{d_mH}\,m
=
-{\ad}^*_{I^{-1}(m)}\,m
\end{equation}
where $\{\,,\,\}$ is the canonical Poisson bracket on $\fg^*$ and ${\ad}^*$ is the
coadjoint action of $\fg$ on $\fg^*$ and $d_mH=I^{-1}(m)$ is the differential of the function
(\ref{EulHam}) (see, e.g.,
\cite{Arn, AK}).

\subsection{The role of central extensions}

In many cases, one is naturally lead to consider central extensions $\widehat{\fg}$ of $\fg$.
A central extension is given by a set of non-trivial 2-\textit{cocycles}
$\mu_i\in{}Z^2(\fg;\bbC)$ with $i=1,\ldots,k$. 
As a vector space,
$\widehat{\fg}\cong\fg\oplus\bbC^k$ where the second summand belongs to the center. 

Consider the dual space
$\widehat{\fg}^*\cong\fg^*\oplus\bbC^k$ 
and fix arbitrary values $(c_1,\ldots,c_k)\in\bbC^k$,
the affine subspace $\fg^*_{(c_1,\ldots,c_k)}\subset\widehat{\fg}^*$ is stable with respect
to the coadjoint action of $\widehat{\fg}$. 
Identifying the affine subspace
$\fg^*_{(c_1,\ldots,c_k)}$ with $\fg^*$ and restricting the coadjoint action of
$\widehat{\fg}$ to $\fg$ (since the center acts trivially), one obtains a $k$-parameter family
of actions of
$\fg$ on $\fg^*$:
\begin{equation}
\label{adDef}
\widehat{\ad}^*_x=
\ad^*_x+\sum_{i=1}^kc_i\,S_i(x)
\qquad\hbox{for}\qquad
x\in\fg.
\end{equation}
where $S_i:\fg\to\fg^*$ are 1-cocycles on $\fg$ with values in the coadjoint representation.
More precisely, the above 1-cocyles $S_i$ are related with the 2-cocycles $\mu_i$ via
$$
\mu_i(x,y)=
\langle
S_i(x),y
\rangle-
\langle
S_i(y),x
\rangle
\qquad\hbox{for}\qquad
(x,y)\in\fg^2.
$$
Formula (\ref{adDef}) is due to Kirillov (see \cite{Kir}) for the details); the
1-cocycles $S_i$ are sometimes called the Souriau cocycles associated with $\mu_i$.

With the above modifications of the coadjoint action and the corresponding Poisson structure,
the Euler equation becomes
\begin{equation}
\label{EulEqDef}
x_t
=
-{\mathrm ad}^*_{I^{-1}(x)}\,x+
c_i\,S_i\left(I^{-1}(x)\right),
\end{equation}
so that one adds to the original equation, which is usually quadratic, some extra
linear terms.

\subsection{The main results}

Our work consists in two parts: 
\begin{itemize}
\item
we introduce an infinite-dimensional Lie algebra
which seems to be a nice generalization of the Virasoro algebra 
with two space variables
(the {\it kinematics});
\item
we find the relevant inertia operators and Euler equations (the {\it dynamics}), in
particular, we are interested in the bi-Hamiltonian Euler equations.
\end{itemize}

We consider the loop algebra over the semidirect product of the Virasoro algebra and its
dual and classify its non-trivial central extensions.
It turns out that one of these central extensions is related simultaneously to the Virasoro
and Kac-Moody algebras. 
We then study the coadjoint representation of this Lie algebra.
We also introduce a Lie superalgebra extending the constructed Lie algebra.
This superalgebra is a generalization of the Neveu-Schwarz algebra. 

We compute several Euler equations corresponding to our Lie algebra.
The first example is very similar to the KP equation (\ref{KP}), yet different
from it. 
In fact, this equation is nothing but KP with supplementary terms and coupled with
another equation.
This equation cannot be reduced to KP,
however, it reduces to KdV.
We do not know if this Hamiltonian system is integrable in any sense.

The second Euler equation we obtained in this paper leads to the following differential
equation
\begin{equation}
\label{OREqn}
g_t
=
g_x\,\partial^{-1}_xg_y-g_y\,g
+c\,\partial^{-1}_x\,g_{yy},
\end{equation}
where $g=g(x,y,t)$ and $c\in\bbC$.
To avoide non-local expressions, one can rewrite this equation as a system:
$$
g_t
+
g_x\,h-h_x\,g
+c\,h_y
=0,
\qquad
g_y+h_x=0.
$$

We prove integrability of (\ref{OREqn}) in the following sense. 
There exists an
infinite hierarchy of vector fields commuting with (\ref{OREqn}) and with commuting flows.
The first commuting fields are: $g_t=g_x$ and $g_t=g_y$; 
one more \textit{higher} field is provided by Example \ref{HerEx}. 
We use the bi-Hamiltonian technique.
More precisely, we obtain equation (\ref{OREqn}) coupled together with another differential
equation (see formula (\ref{VNProPE})) so that the system of two equations is a
bi-Hamiltonian vector field.

Equation (\ref{OREqn}) was studied in \cite{FK} (see formula (33)) and \cite{FK1}. It has
also been considered in differential geometry \cite{Dun}. In a more general setting, this
equation is a the second term of the so-called universal hierarchy \cite{MS}.

Although equation (\ref{OREqn}) resembles to KP and especially to BKP (\ref{BKP}), it is
different: there are no cubic terms and, foremost, the
sign in the quadratic term is different. This may be important, especially if one works over
$\bbR$ (rather than over $\bbC$).


\subsection{Historical overview}

The most classic case is related to the Lie group $G=\SO(3)$ and
the inertia operator $I$
given by a symmetric tensor in $S^2\fg$ (the usual inertia tensor on the rigid solid). One
gets the genuine Euler equation.

The first generalization was obtained by V.I. Arnold (1966) for the hydrodynamical Euler
equation of an incompressible fluid, hence the name of \textit{Euler-Arnold} is sometimes
granted to these equations. The relevant group in this case is the group $\SDiff(D)$ of
volume-preserving diffeomorphisms of a domain $D\subset\bbR^2$ or $\bbR^3$ (see the books
\cite{Arn,AK}).

Some interesting examples, such as the Landau-Lifchitz
equation, correspond to the Euler equations on the Kac-Moody groups (see \cite{Kir1}).

Another example has already been mentioned: the KdV equation is an Euler equation on the
Virasoro-Bott group (see \cite{KO}). This group is defined as the unique (up to
isomorphism) non-trivial central extension of the group $\Diff(S^1)$ of all diffeomorphisms of
$S^1$. The inertia operator is given by the standard $L^2$-metric on $S^1$.

In \cite{Mis} and \cite{KM} different choices of the metric on
$\Diff(S^1)$ (and thus of the inertia operator) were considered in order to get some other
equations than KdV, such that the Hunter-Saxon and the Camassa-Holm equations. 
Let us also mention that there is a huge literature containing different generalizations of
KdV (in the super case, matrix versions, etc.) in the dimension $1+1$.
All these generalizations are related to some extensions of the Virasoro algebra.

Let us finally stress that the property of an evolution equation to be an Euler equation
associated with some Lie group (or Lie algebra) is important for the following reason:
it allows one to deal with the equation as with the geodesic flow and to apply a wide spectrum
of methods specific for differential geometry (see, e.g., \cite{Mis1}).

\section{The Virasoro algebra and its loop algebra}

In this section we introduce the preliminary examples of infinite-dimensional Lie algebras
that we will consider. 
We define the Virasoro algebra and show how to obtain the KdV
equation as an Euler equation on it. 
We then consider the loop group
$L\left(\Diff(S^1)\right)$.  
We classify  non-trivial central extensions of the
corresponding Lie algebra.

\subsection{Reminder: the Virasoro algebra and KdV equation}\label{VKdVSec}

The \textit{Virasoro algebra} is defined as the central extension of $\Vect(S^1)$ given by the
2-cocycle
\begin{equation}
\label{GFCoc}
\mu\Big(
f(x)\,\frac{\partial}{\partial x},\,g(x)\,\frac{\partial}{\partial x}
\Big)=
\int_{S^1}f\,g_{xxx}\,dx.
\end{equation}
This cocycle was found in \cite{GF} and is known as the \textit{Gelfand-Fuchs cocycle}.
The cohomology group $H^2(\Vect(S^1);\bbC)$ is one-dimensional so that the cocycle
(\ref{GFCoc}) defines the unique (up to isomorphism) non-trivial central extension of
$\Vect(S^1)$.

Consider a natural family of modules over $\Vect(S^1)$ (and therefore over the Virasoro
algebra with the trivial action of the center). 
Let $\cF_\l$ be the space of $\l$-\textit{densities}
(or the space of tensor densities of degree $\l$) on $S^1$
$$
\cF_\l=
\left\{
\varphi(x)\,dx^\l
\left|\,
\varphi(x)\in{}C^\infty(S^1)
\right.
\right\},
$$
where $\l\in\bbC$.
The action of $\Vect(S^1)$ on the space $\cF_\l$ is given by the first-order
differential operator
\begin{equation}
\label{LDTDFor}
L_f^\l
\left(
\varphi(x)\,dx^\l
\right)
=\left(
f\,\varphi_x+
\l\,f_x\,\varphi
\right)dx^\l
\end{equation}
which is nothing but the Lie derivative along the vector field
$f=f(x)\,\frac{\partial}{\partial x}$.

Let us calculate the coadjoint action of the Virasoro algebra.
The dual space, $\Vect(S^1)^*$, corresponds to the space of all distributions on $S^1$.
Following \cite{Kir1}, we will consider only the
\textit{regular part} of this dual space that consists of differentiable 2-densities, that is
$
\Vect(S^1)_{\rm reg}^*=\cF_2
$
with the natural pairing
$$
\Big\langle
f(x)\,\frac{\partial}{\partial x},\,
u(x)\,dx^2
\Big\rangle=
\int_{S^1}f(x)\,u(x)\,dx.
$$
The coadjoint action of $\Vect(S^1)$ coincides with the $\Vect(S^1)$-action on $\cF_2$.
The Souriau cocycle on $\Vect(S^1)$ corresponding to the Gelfand-Fuchs cocycle is
$$
S\Big(f(x)\,\frac{\partial}{\partial x}\Big)
=f_{xxx}\,dx^2
$$
and one finally obtains the
coadjoint action of the Virasoro algebra:
\begin{equation}
\label{CoacVir}
\widehat{\ad}^*_{f(x)\,\frac{\partial}{\partial x}}
\left(
u(x)\,dx^2
\right)=
\left(
f\,u_x+2\,f_x\,u
+
c\,f_{xxx}\right)
dx^2.
\end{equation}

Consider the simplest quadratic Hamiltonian function on $\Vect(S^1)^*$
$$
H\left(u(x)\,dx^2\right)=\frac{1}{2}\,\int_{S^1}u(x)^2\,dx
$$
corresponding to the inertia operator 
$I(f(x)\,\frac{\partial}{\partial x})=f(x)\,dx^2$.
The following result was obtained in \cite{KO}.

\begin{prop}
\label{KdVFirstProp}
The Euler equation on the Virasoro algebra corresponding to the Hamiltonian $H$ is precisely
the KdV equation (\ref{KdV}).
\end{prop}

\begin{proof}
Immediately follows from formul{\ae} (\ref{EulEq}) and (\ref{CoacVir}).
\end{proof}

For more details about the Virasoro algebra,
its modules and its cohomology see \cite{Fuk} and \cite{GR}.

\subsection{The loop group on $\Diff(S^1)$ and the loop algebra on $\Vect(S^1)$}

We wish to extend the Virasoro algebra to the case of two space variables.
A natural way to do this is to consider the loops on it.

One defines the loop group on $\Diff(S^1)$ as follows
$$
L\left(\Diff(S^1)\right)=
\left\{
\vp:S^1\to\Diff(S^1)\,
|\quad\vp
\quad\hbox{is differentiable}
\right\},
$$
the group law being given by
$$
\left(\vp\circ\psi\right)(y)=
\vp(y)\circ\psi(y),
\qquad
y\in{}S^1.
$$

\begin{rmk}
{\rm
Let us stress that there are no difficulties in defining differentiable maps with values in
$\Diff(S^1)$. Indeed, $L\left(\Diff(S^1)\right)$ is naturally embedded into the space of $C^\infty$-maps
on $S^1\times{}S^1$ with values in $S^1$.
}
\end{rmk}

In the similar way, we construct the Lie algebra $L\left(\Vect(S^1)\right)$ 
consisting of vector fields
on $S^1$ depending on one more independent variable $y\in{}S^1$. 
The loop variable is thus denoted by $y$ and the variable on the ``target'' copy of $S^1$ by
$x$. The elements of
$L\left(\Vect(S^1)\right)$ are of the form: $f(x,y)\frac{\partial}{\partial x}$ where
$f\in{}C^\infty(S^1\times{}S^1)$ and the Lie bracket reads as follows
$$
\left[
f(x,y)\frac{\partial}{\partial x},g(x,y)\frac{\partial}{\partial x}
\right]=
\left(
f(x,y)\,g_x(x,y)-
f_x(x,y)\,g(x,y)
\right)\frac{\partial}{\partial x}.
$$
It is easy to convince oneself that $L\left(\Vect(S^1)\right)$ is the Lie algebra of 
$L\left(\Diff(S^1)\right)$ in
the usual weak sense for the infinite-dimensional case; a one-parameter group argumentation
gives an identification between the tangent space to $L\left(\Diff(S^1)\right)$ at the
identity and $L\left(\Vect(S^1)\right)$, equipped with its Lie bracket.

We will now classify non-trivial central extensions of the Lie algebra
$L\left(\Vect(S^1)\right)$ and therefore calculate $H^2(L\left(\Vect(S^1)\right);\bbC)$. 
This result can be deduced from a more general one
that we will need later. 

\subsection{Central extensions of tensor products}

Following the work of Zusmanovich \cite{Zus}, one can calculate the
cohomology group $H^2(\fg\otimes\cA)$ for a Lie algebra $\fg$ and a commutative algebra $\cA$
over a field $k$, the Lie bracket on $\fg\otimes\cA$ being defined by
$$
\left[
x_1\otimes{}a_1,x_2\otimes{}a_2
\right]=
\left[
x_1,x_2
\right]\otimes{}a_1a_2,
\qquad{}x_i\in\fg,
\;{}a_i\in\cA,
\quad{}i=1,2.
$$
From the results of \cite{Zus}, one can easily deduce the following

\begin{prop}
\label{ProHom}
If $\fg=[\fg,\fg]$, then
$$
H^2(\fg\otimes\cA;k)=
\left(
H^2(\fg;k)\otimes\cA^\prime
\right)
\oplus
\left(
\Inv_\fg{}S^2(\fg^*)\otimes{}HC^1(\cA)
\right)
$$
where $HC^1(\cA)$ is the first group of cyclic cohomology of the $k$-algebra $\cA$ and
$\Inv_\fg{}S^2(\fg^*)$ is the space of $\fg$-invariant symmetric bilinear maps from $\fg$ into
$k$, while $\cA^\prime=\Hom_k(\cA,k)$ represents the dual of $\cA$.
\end{prop}

One can give explicit formul{\ae} for the cohomology classes.

\begin{itemize}
\item
Given $\mu\in{}Z^2(\fg;k)$ and $\l\in\cA^\prime$, one gets
$\mu_\l\in{}Z^2(\fg\otimes\cA;k)$ defined by
\begin{equation}
\label{CocOneEq}
\mu_\l(x_1\otimes{a_1},x_2\otimes{a_2})=
\mu(x_1,x_2)\,\l(a_1a_2).
\end{equation}
\item
Given $K\in\Inv_\fg{}S^2(\fg^*)$ and $\ell\in{}HC^1(\cA)$, one gets
$K_\ell\in{}Z^2(\fg\otimes\cA;k)$ defined by
\begin{equation}
\label{CocTwoEq}
K_\ell(x_1\otimes{a_1},x_2\otimes{a_2})=
K(x_1,x_2)\,\ell(a_1da_2),
\end{equation}
where $d$ is the K\"ahler derivative.
\end{itemize}
For the general results on cyclic homology and cohomology see \cite{Lod}.

\begin{exe}
{\rm
If $\fg$ is a finite-dimensional semisimple Lie algebra and $\cA=C^\infty(S^1)$, formula
(\ref{CocTwoEq}) defines the Kac-Moody cocycle
$$
KM(x_1\otimes{a_1},x_2\otimes{a_2})=
K(x_1,x_2)\,\int_{S^1}a_1da_2,
$$
where $K$ is the Killing form. 
}
\end{exe}

\begin{rmk}
{\rm
In the general situation, one can call the cohomology classes of the cocycles
(\ref{CocTwoEq}) the classes  ``of Kac-Moody type''. For instance, such a class on the loop
algebra over the algebra of pseudodifferential symbols has already been used in \cite{RS} in
order to obtain the KP equation as a Hamiltonian system.
}
\end{rmk}

\subsection{Central extensions of $L\left(\Vect(S^1)\right)$}

In our case, $\fg=\Vect(S^1)$ and $\cA=C^\infty(S^1)$, one has the following well-known
statement:
$\Inv_\fg{}S^2(\fg^*)=0$, that is, there is no invariant bilinear symmetric form (``Killing
form'') on $\Vect(S^1)$ (see, e.g., \cite{Fuk}).
Proposition \ref{ProHom} then implies the following result.
\begin{prop}
\label{ProNeshHom}
One has
$$
H^2(L\left(Vect(S^1)\right);\bbC)=
C^\infty(S^1)^\prime
$$
To a distribution $\l\in{}C^\infty(S^1)^\prime$ one associates a 2-cocycle
$\mu_\l\in{}Z^2(L\left(\Vect(S^1)\right);\bbC)$ given by formula (\ref{CocOneEq}) with $\mu$
being the Gelfand-Fuchs cocycle (\ref{GFCoc}).
\end{prop}
Similar results were obtained in \cite{RSS} in a slightly different context. 

Proposition \ref{ProNeshHom} provides a classification of non-trivial central extensions of
the loop algebra $L\left(\Vect(S^1)\right)$. 
This is rather a ``negative result'' for us since it implies that all these central extensions
are of Virasoro type. 
The KP equation (\ref{KP}) contains two different central
charges, $c_1$ and $c_2$, and the second one does not belong to the Virasoro type but 
to the Kac-Moody one. 
It is clear then that KP-type (and BKP-type) equations cannot be
obtained as Euler equations associated with the group $L\left(\Diff(S^1)\right)$. 
One therefore needs to introduce another group with richer second cohomology.

We will discuss further
generalizations of $L\left(\Vect(S^1)\right)$ in the Appendix.
However, this Lie algebra is not the one we are interested in.

\section{The cotangent Virasoro algebra and its loop algebra}

In this section we introduce the main object of our study. 
We consider the Lie algebra of loops
associated with the cotangent Lie algebra $T^*\Vect(S^1)$ and calculate its central
extensions. 
Unlike the loop Lie algebra $L\left(\Vect(S^1)\right)$, the constructed Lie
algebra  simultaneously has non-trivial central extensions of the Virasoro and the Kac-Moody
types.

\subsection{General setting: the cotangent group and its Lie algebra}

The cotangent space $T^*G$ of a Lie group $G$ is naturally identified with
the semi-direct product $G\ltimes\fg^*$ where $G$ acts on $\fg^*$ by the coadjoint action
$\Ad^*$. The space
$T^*G$ is then a Lie group with the product
$$
(g_1,u_1)\cdot(g_2,u_2)=
(g_1g_2,\,u_1+\Ad^*_{g_1}u_2).
$$
The corresponding Lie algebra $T^*\fg$ is the semi-direct product $\fg\ltimes\fg^*$ equipped
with the commutator
\begin{equation}
\label{CmtLATilg}
\left[
(x_1,u_1),(x_2,u_2)
\right]=
\left(
[x_1,x_2],\,\ad^*_{x_1}u_2-\ad^*_{x_2}u_1
\right).
\end{equation}
The evaluation map gives a natural symmetric bilinear form on $T^*\fg$, namely
\begin{equation}
\label{TwoFormFor}
K\left((x_1,u_1),(x_2,u_2)\right)=
\langle{}u_1,x_2\rangle+
\langle{}u_2,x_1\rangle.
\end{equation}
Furthermore, this form is non-degenerate so that one has a Killing type form on this Lie
algebra.

\begin{rmk}
{\rm
Let us mention that the semi-direct product $\fg\ltimes\fg^*$ is the simplest case of
so-called Drinfeld's double which corresponds to the trivial Lie-Poisson structure.
}
\end{rmk}

\subsection{The cotangent loop group and algebra}

We will consider the cotangent group $G=T^*\Diff(S^1)$ and we will be particularly interested
in the associated loop group. We will use the notation
$\widetilde{G}=L\left(T^*\Diff(S^1)\right)$ for short.
One has
$$
\widetilde{G}=
L\left(
\Diff(S^1)\ltimes\cF_2
\right)=
L\left(
\Diff(S^1)
\right)
\ltimes{}L(\cF_2).
$$
Consider the semidirect product
$$
\fg=\Vect(S^1)\ltimes\cF_2.
$$
The Lie algebra corresponding to the group $\widetilde{G}$ is
\begin{equation}
\label{DefLATilg}
\widetilde{\fg}=
L(\Vect(S^1)\ltimes\cF_2)=
L\left(\Vect(S^1)\right)\ltimes{}L(\cF_2).
\end{equation}
An element of $\widetilde{\fg}$ is a couple $(f,u)$ where $f$ and $u$ are
$C^\infty$-tensor fields on $S^1\times{}S^1$ of the following form
$$
(f,u)=
f(x,y)\,\frac{\partial}{\partial x}+
u(x,y)\,dx^2.
$$
The commutator (Lie bracket) is defined accordingly to (\ref{CmtLATilg}).

\begin{rmk}
{\rm
It is easy to check that, with the right convention on the duality, one has
$L(\cF_2)=L\left(\Vect(S^1)^*\right)=L\left(\Vect(S^1)\right)^*$; the choosen form
$L(\cF_2)$ will be more suitable for our computations.
}
\end{rmk}

\subsection{Central extensions of $\widetilde{\fg}$}

Let us calculate the cohomology space
$H^2(\widetilde{\fg},\bbC)$ using Proposition \ref{ProHom}. 
The following result is a
classification of central extensions of the Lie algebra
$\widetilde{\fg}$.

\begin{thm}
\label{ProKhor}
One has
$$
H^2(\widetilde{\fg};\bbC)=
C^\infty(S^1)^\prime\oplus\bbC.
$$
\end{thm}

\begin{proof}
From the
classical results on the cohomology of the Virasoro algebra and its representations, one has
$H^2(\fg,\bbC)=\bbC$ where, as above, $\fg=\Vect(S^1)\ltimes\cF_2$, cf. \cite{Fuk} and
\cite{GF}. The non-trivial cohomology class is again generated by the Gelfand-Fuchs
cocycle. More precisely,
$$
\m^\prime\left((f,u),(g,v)\right)=\m(f,g),
$$
where $\m$ is as in (\ref{GFCoc}).

Furthermore, one has $\Inv_\fg{}S^2(\fg^*)=\bbC$, where the generator is provided by the
2-form (\ref{TwoFormFor}) and it is easy to check that there are no other generators.
Finally, $HC^1(\cA)=\bbC$ and is generated by the volume form (see, e.g., \cite{Lod}).
\end{proof}

Let us give the explicit formul{\ae} of non-trivial 2-cocycles on $\widetilde{\fg}$.
A distribution $\l\in{}C^\infty(S^1)^\prime$ corresponds to a 2-cocycle of the first class
(\ref{CocOneEq}) given by
$$
\m_\l\left((f,u),(g,v)\right)=
\l\Big(\int_{S^1}f\,g_{xxx}\,dx\Big),
$$
these are the Virasoro type extensions.
For the particular case where $\l(a(y))=\int_{S^1}a(y)dy$, such a 2-cocycle will be denoted by
$\m_1$ so that one has
\begin{equation}
\label{GFCocModifFor}
\m_1\left((f,u),(g,v)\right)=
\int_{S^1\times{}S^1}f\,g_{xxx}\,dxdy.
\end{equation}
Another non-trivial cohomology class is provided by the 2-cocycle
\begin{equation}
\label{CMCocModifFor}
\m_2\left((f,u),(g,v)\right)=
\int_{S^1\times{}S^1}
\left(
f\,v_{y}-g\,u_y
\right)\,dxdy
\end{equation}
of Kac-Moody type (\ref{CocTwoEq}).

\subsection{The Lie algebra $\widehat{\fg}$}

We define the Lie algebra $\widehat{\fg}$ as the two-dimensional central
extension of $\widetilde{\fg}$ given by the cocycles $\m_1$ and $\m_2$.
As a vector space,
$$
\widehat{\fg}=
\widetilde{\fg}\oplus\bbC^2,
$$
where the summand $\bbC^2$ is the center of $\widehat{\fg}$.
The commutator in $\widehat{\fg}$ is given by the following explicit expression
which readily follows from the above formul{\ae}.
\begin{equation}
\label{CommutHatEq}
\begin{array}{rcl}
\displaystyle
\left[f\,\frac{\partial}{\partial x}+u\,dx^2,\,
g\,\frac{\partial}{\partial x}+v\,dx^2\right]
&=&
\displaystyle
\left(f\,g_x-f_x\,g\right)\frac{\partial}{\partial x}\\[12pt]
&&
\displaystyle
+\left(f\,v_x+2\,f_x\,v-g\,u_x-2\,g_x\,u\right)dx^2\\[12pt]
&&
\displaystyle
+
\Big(\int_{S^1\times{}S^1}f\,g_{xxx}\,dxdy\,,
\int_{S^1\times{}S^1}
\left(
f\,v_{y}-g\,u_y
\right)\,dxdy
\Big),
\end{array}
\end{equation}
where the last term is an element of the center of $\widehat{\fg}$.
(Note that we did not write the central elements in the left hand side since they do not
enter into the commutator).

The Lie algebra $\widehat{\fg}$ and its coadjoint representation is the main object of our
study. This algebra is a natural two-dimensional generalization of the Virasoro algebra.
Further generalizations will be described in the Appendix.

\section{The coadjoint representation of $\widehat{\fg}$}

According to the general viewpoint of symplectic geometry and mechanics,
the coadjoint representation of a Lie algebra plays a very special role
for all sorts of applications.

It was observed by Kirillov \cite{Kir1} and Segal \cite{Seg} that the
coadjoint action of the Virasoro group and algebra coincides with
the natural action of $\Diff(S^1)$ and $\Vect(S^1)$, respectively, on the space of
Sturm-Liouville operators. The Casimir functions (i.e., the invariants of the coadjoint
action) are then expressed in terms of the monodromy operator. 

In this section we recall the Kirillov-Segal result and generalize it to the case of the Lie
algebra $\widehat{\fg}$.

\subsection{Computing the coadjoint action}

Let us start with the explicit expression for the coadjoint action of $\widehat{\fg}$.

As usual, in the case of semidirect product of a Lie algebra and its dual,
the (regular) dual space $\widetilde{\fg}^*$ is identified with $\widetilde{\fg}$;
the natural pairing being given by
\begin{equation}
\label{QF}
\left\langle
f\,\frac{\partial}{\partial x}+u\,dx^2,\,
g\,\frac{\partial}{\partial x}+v\,dx^2
\right\rangle=
\int_{S^1\times{}S^1}
\left(
f\,v+g\,u
\right)dxdy
\end{equation}
which is nothing but a specialization of the form (\ref{TwoFormFor}).
Furthermore, one immediately obtains the Souriau 1-cocycles on $\widetilde{\fg}$
\begin{equation}
\label{SouEFor}
\begin{array}{lll}
\displaystyle
S_1\Big(f\,\frac{\partial}{\partial x}+u\,dx^2\Big)&=&
\displaystyle
f_{xxx}\,dx^2,\\[12pt]
\displaystyle
S_2\Big(f\,\frac{\partial}{\partial x}+u\,dx^2\Big)&=&
\displaystyle
f_y\,\frac{\partial}{\partial x}+u_y\,dx^2
\end{array}
\end{equation}
corresponding to $\m_1$ and
$\m_2$, respectively.

We are ready to give the expression of the coadjoint action of the ``extended'' Lie algebra 
$\widehat{\fg}$ on the regular dual space $\widetilde{\fg}^*\cong\widetilde{\fg}$.

\begin{prop}
\label{CPp}
The coadjoint action of the Lie algebra $\widehat{\fg}$ is given by
\begin{equation}
\label{CExtnEFor}
\begin{array}{rcl}
\displaystyle
\widehat{\ad}^*_{\left(
f,u
\right)}\,
\Big(
g\,\frac{\partial}{\partial x}+v\,dx^2
\Big)
&=&
\displaystyle
\left(f\,g_x-f_x\,g+c_2\,f_y\right)\frac{\partial}{\partial x}\\[12pt]
&&
\displaystyle
+\left(f\,v_x+2\,f_x\,v-u_x\,g-2\,u\,g_x+c_1\,f_{xxx}+c_2\,u_y\right)dx^2,
\end{array}
\end{equation}
while the center acts trivially.
\end{prop}
\begin{proof}
According to formula (\ref{adDef}), one has to calculate the coadjoint action of
$\widetilde{\fg}$, which coincides with the adjoint one (cf. formula (\ref{CommutHatEq}) with
no central terms) and then to add the Souriau cocycles.
\end{proof}

Our next task is to investigate a ``geometric meaning'' of the coadjoint action
(\ref{CExtnEFor}).
However, this result will not be relevant for our main task: computing the Euler equations on
$\widehat{\fg}$.
Sections \ref{KSect}--\ref{SuSect} can be omitted at the first reading.

\subsection{The Virasoro coadjoint action and Sturm-Liouville operators}
\label{KSect}

Let us recall the Kirillov-Segal result on the coadjoint action of the Virasoro algebra
(cf. \cite{Kir1,Seg,OT}). Consider the space of second-order linear differential
operators
\begin{equation}
\label{SLInterFor}
A=
2c\left(\frac{d}{d x}\right)^2
+u(x),
\end{equation}
where $u(x)\in{}C^\infty(S^1)$.
This is a Sturm-Liouville operator with periodic potential (also called a Hill operator).
Define an action of $\Vect(S^1)$ on this space by the formula
\begin{equation}
\label{SLActFor}
\cL_f(A):=
L_f^{\frac{3}{2}}\circ{}A-
A\circ{}L_f^{-\frac{1}{2}},
\end{equation}
where $f=f(x)\,\frac{d}{d x}$ and where $L^\l_f$ is the first-order differential operator
$$
L^\l_f=f\,\frac{\partial}{\partial x}+\l\,f_x
$$
corresponding to the Lie derivative on the
space $\cF_\l$ of $\l$-densities (\ref{LDTDFor}).

An elementary computation shows that the result of the action (\ref{SLActFor}) is a
differential operator of order 0 (that is, an operator of multiplication by a function). 
More precisely, one has
$$
\cL_f(A)=
f\,v_x+2\,f_x\,v
+
c\,f_{xxx}.
$$
This formula coincides with formula (\ref{CoacVir}) of the coadjoint action of the
Virasoro algebra.
We obtained the following result:
\textit{the space of operators (\ref{SLInterFor}) equipped with the $\Vect(S^1)$-action is
isomorphic to the coadjoint representation of the Virasoro algebra}.

\begin{rmk}
{\rm
The operator $A$ is understood as a differential operator on tensor
densities on $S^1$, namely $A:\cF_{-\frac{1}{2}}\to\cF_{\frac{3}{2}}$. The
expression (\ref{LDTDFor}) is the Lie derivative on the space $\cF_\l$ and thus the action
(\ref{SLActFor}) is natural. This geometric meaning of the Sturm-Liouville operator has
already been known by classics. 
}
\end{rmk}

This remarkable coincidence is just a simple observation, but it has important consequences.
It relates the Virasoro algebra with the projective differential geometry
(we refer to \cite{OT} and references therein for more details).
For instance, one can now interpret the monodromy operator associated with
$A$ as an invariant of the coadjoint action and eventually obtain a classification of the
coadjoint orbits of the Virasoro algebra.
Sturm-Liouville operators are also closely related to the KdV hierarchy.

Let us mention that another geometric approach to the study of the Virasoro coadjoint orbits
can be found in
\cite{BFP}.

\subsection{Relations with the Neveu-Schwarz superalgebra}

It order to understand the origin of the Sturm-Liouville operator in the context of Virasoro
algebra, we will apply Kirillov's sophisticated method \cite{Kir}
that uses \textit{Lie superalgebras}.
Let us note that this is quite an unsusual situation when superalgebra helps to better
understand the susual (``non-super'') situation.

Let us recall the definition of the \textit{Neveu-Schwarz} Lie superalgebra.
Consider the direct sum
$$
\mathfrak{k}=\Vect(S^1)\oplus\cF_{-\frac{1}{2}}
$$
and define the structure of a Lie superalgebra on the space $\mathfrak{k}$ by
$$
\begin{array}{rcl}
\displaystyle
\left[
f\,\frac{\partial}{\partial x}+\vp\,dx^{-\frac{1}{2}},\,
g\,\frac{\partial}{\partial x}+\psi\,dx^{-\frac{1}{2}}
\right]
&=&
\displaystyle
\left(
fg_x-f_xg+\vp\psi
\right)\frac{\partial}{\partial x}\\[12pt]
&&
\displaystyle
+\Big(
f\psi_x-\frac{1}{2}\,f_x\psi-g\vp_x+\frac{1}{2}\,g_x\vp
\Big)\,dx^{-\frac{1}{2}},
\end{array}
$$
which is symmetric on the odd part $\mathfrak{k}_1=\cF_{-\frac{1}{2}}$.
The Jacobi (super)identity can be easily checked.
Furthermore, the Gelfand-Fuchs cocycle (\ref{GFCoc}) can be extended to the superalgebra
$\mathfrak{k}$:
\begin{equation}
\label{MCocy}
\mu
\Big(
f\,\frac{\partial}{\partial x}+\vp\,dx^{-\frac{1}{2}},\,
g\,\frac{\partial}{\partial x}+\psi\,dx^{-\frac{1}{2}}
\Big)
=
\int_{S^1}
\left(
fg_{xxx}+2\,\vp\psi_{xx}
\right)dx.
\end{equation}
This 2-cocycle defines a central extension $\widehat{\mathfrak{k}}$ of
$\mathfrak{k}$ called the Neveu-Schwarz algebra.

The dual space of this superalgebra is
$$
\widehat{\mathfrak{k}}^*=\cF_2\oplus\cF_{\frac{3}{2}},
$$
since $\cF_{-\frac{1}{2}}^*\cong\cF_{\frac{3}{2}}$.
The coadjoint action of $\widehat{\mathfrak{k}}$ can be easily calculated:
\begin{equation}
\label{CAk}
\begin{array}{rcl}
\displaystyle
\widehat{\ad}^*_{f\,\frac{\partial}{\partial x}+\vp\,dx^{-\frac{1}{2}}}
\left(
u\,dx^2+\a\,dx^\frac{3}{2}
\right)
&=&
\displaystyle
\Big(
f\,v_x+2\,f_x\,v
+
c\,f_{xxx}+\frac{1}{2}\,\vp\,\a_x+\frac{3}{2}\vp_x\,\a
\Big)\,dx^2\\[12pt]
&&
\displaystyle
\Big(
f\,\a_x+\frac{3}{2}\,f_x\,\a
+u\,\vp+
2\,c\,\vp_{xx}
\Big)\,dx^\frac{3}{2}.
\end{array}
\end{equation}
The operator (\ref{SLInterFor}) is already present in this formula:
it gives the action of the \textit{odd} part of 
$\widehat{\mathfrak{k}}$ on
the \textit{even} part of  $\widehat{\mathfrak{k}}^*$, namely
$$
\widehat{\ad}^*_{\vp\,dx^{-\frac{1}{2}}}
\left(u\,dx^2\right)=A(\vp),
$$
where $A$ is as in (\ref{SLInterFor}).

This way, the Sturm-Liouville operator naturally appears in the Virasoro context.
The Kirillov-Segal result can now be deduced simply from the Jacobi identity for the
superalgebra
$\widehat{\mathfrak{k}}$.

\subsection{Coadjoint action of $\widehat{\fg}$ and matrix differential operators}

It turns out that the coadjoint action of $\widehat{\fg}$ given by the formula
(\ref{CExtnEFor}) can also be realized as an action of the non-extended algebra
$\widetilde{\fg}$ on some space of differential operators.

We introduce the space of $2\times2$-matrix differential operators
\begin{equation}
\label{MatSLEq}
A=
\left(
\begin{array}{cc}
\displaystyle
-c_2\,\frac{\partial}{\partial y}
+g\,\frac{\partial}{\partial x}
-\frac{1}{2}\,g_x
&
\displaystyle
0\\[14pt]
\displaystyle
2c_1\left(\frac{\partial}{\partial x}\right)^2
+v
&
\displaystyle
-c_2\,\frac{\partial}{\partial y}
+g\,\frac{\partial}{\partial x}+
\frac{3}{2}\,g_x
\end{array}
\right)
\end{equation}
where $g$ and $v$ are functions in $(x,y)$, that is $g,v\in{}C^\infty(S^1\times{}S^1)$.
Let us define an action of the Lie algebra $\widetilde{\fg}$ on the space of operators
(\ref{MatSLEq}).

Let us now denote by $\widetilde{\cF}_\l$ the space of loops in $\cF_\l$, i.e.,
of tensor fields on $S^1\times{}S^1$ of the form
$
\vp=\vp(x,y)\,dx^\l.
$
The direct sum $\widetilde{\cF}_{-\frac{1}{2}}\oplus\widetilde{\cF}_{\frac{3}{2}}$ is
equipped with a $\widetilde{\fg}$-action
\begin{equation}
\label{TenAct}
L_{
f\,\frac{\partial}{\partial x}+u\,dx^2}
\left(
\begin{array}{l}
\vp\,dx^{-\frac{1}{2}}\\[10pt]
\psi\,dx^{\frac{3}{2}}
\end{array}
\right)
=
\left(
\begin{array}{l}
\left(
f\,\vp_x-\frac{1}{2}\,f_x\,\vp
\right)dx^{-\frac{1}{2}}\\[10pt]
\left(
f\,\psi_x+\frac{3}{2}\,f_x\,\psi+u\,\vp
\right)dx^{\frac{3}{2}}
\end{array}
\right)
\end{equation}

Assume the operator $A$ acting on the above space of tensor densities:
$$
A:\widetilde{\cF}_{-\frac{1}{2}}\oplus\widetilde{\cF}_{\frac{3}{2}}\to
\widetilde{\cF}_{-\frac{1}{2}}\oplus\widetilde{\cF}_{\frac{3}{2}}
$$
Then the $\widetilde{\fg}$-action is naturally defined by the commutator
with the action (\ref{TenAct}):
\begin{equation}
\label{TildAct}
\cL_{f\,\frac{\partial}{\partial x}+u\,dx^2}(A):=
\left[
L_{f\,\frac{\partial}{\partial x}+u\,dx^2},\,A
\right].
\end{equation}
The following statement is a generalization of the Kirillov-Segal result.

\begin{thm}
\label{KSGenThm}
The action (\ref{TildAct}) of $\widetilde{\fg}$ on the space of differential operators
(\ref{MatSLEq}) coincides with the coadjoint action (\ref{CExtnEFor})
of the Lie algebra $\widehat{\fg}$.
\end{thm}
\begin{proof}
This statement can be checked by a straightforward computation.
\end{proof}

In Section \ref{SuSect} we will give another, conceptual proof of this theorem.
Let us also mention that similar results were obtained in \cite{MOR,Mar} for different
examples of Lie algebras generalizing Virasoro.

Theorem \ref{KSGenThm} implies that the invariants of the operators (\ref{MatSLEq}) are now
invariants of the coadjoint orbits. It would be interesting to investigate these invariants,
for instance, if there is an analogue of the monodromy operator of $A$.

\subsection{Poisson bracket of tensor densities}
\label{PBTDSec}

Many of the explicit formul{\ae} that we calculated in this paper (see for instance,
(\ref{CommutHatEq}), (\ref{TenAct}) and (\ref{CAk})) use the  bilinear operation on tensor
densities
$$
\widetilde{\cF}_\l\otimes\widetilde{\cF}_\mu\to\widetilde{\cF}_{\l+\mu+1}
$$
given by
\begin{equation}
\label{SchouBt}
\left\{
\vp\,dx^\l,\,\psi\,dx^\mu
\right\}
=
\left(
\l\,\vp\,\psi_x-\mu\,\vp_x\,\psi
\right)dx^{\l+\m+1}.
\end{equation}
It is easy to check that the bilinear maps (\ref{SchouBt}) are invariant.
We will call the operation  (\ref{SchouBt}) the \textit{Poisson bracket} of tensor densities.

Let us rewrite some of the main formul{\ae} using the Poisson
bracket. The commutator in 
$\widetilde{\fg}=\widetilde{\cF}_{-1}\oplus\widetilde{\cF}_2$ is
simply
\begin{equation}
\label{InvComF}
\left[
(f,u),\,(g,v)
\right]
=
\left(
\{f,\,g\},\,\{f,\,v\}-\{g,\,u\}
\right)
\end{equation}
while the formula (\ref{TenAct}) of $\widetilde{\fg}$-action on
$\widetilde{\cF}_{-\frac{1}{2}}\oplus\widetilde{\cF}_{\frac{3}{2}}$ reads in invariant terms
as follows
\begin{equation}
\label{InvActF}
L_{(f,\,u)}
\left(
\begin{array}{l}
\vp\\[8pt]
\psi
\end{array}
\right)
=
\left(
\begin{array}{l}
\{f\,\vp\}\\[8pt]
\{f\,\psi\}+u\,\vp
\end{array}
\right).
\end{equation}

\subsection{A Lie superalgebra extending $\widehat{\fg}$}
\label{SuSect}

Although the proof of the Theorem \ref{KSGenThm} is, indeed, straighforward, it does not
clarify the origin of the operators (\ref{MatSLEq}). 
The Kirillov method using the Lie superalgebras proved to be
universal (see \cite{Ovs} for the details) and will be useful in our case.
The Lie superalgebra we define in this section generalizes the Neveu-Schwarz algebra in the
same sense as $\widehat{\fg}$ generalizes the Virasoro algebra. 
It can be called the looped cotangent Neveu-Schwarz algebra.

Consider the $\widetilde{\fg}$-module
$\widetilde{\cF}_{-\frac{1}{2}}\oplus\widetilde{\cF}_{\frac{3}{2}}$ and denote it by
$\widetilde{\fg}_1$.
We will define the Lie superalgebra structure on
$$
\widetilde{\fG}=\widetilde{\fg}\oplus\widetilde{\fg}_1.
$$
Since we already know
$\widetilde{\fg}$-action (\ref{TenAct}) on
$\widetilde{\fg}_1$, it remains to define the symmetric
operation
$$
\widetilde{\fg}_1\otimes\widetilde{\fg}_1\to\widetilde{\fg},
$$
i.e., the ``anticommutator''.
Similarly to (\ref{InvComF}) and (\ref{InvActF}), let us set
\begin{equation}
\label{InvAntComF}
\left[
(\vp,\a),\,(\psi,\b)
\right]
=
\left(
\vp\,\psi,\,\{\vp,\,\b\}+\{\psi,\,\a\}
\right).
\end{equation}
One immediately obtains the following

\begin{prop}
\label{SAProp}
Formula (\ref{InvAntComF}) defineds a structure of a Lie superalgebra
on $\widetilde{\fG}$.
\end{prop}

The cocycles (\ref{GFCocModifFor}) and (\ref{CMCocModifFor}) can be extended from
$\widetilde{\fg}$ to $\widetilde{\fG}$ as even cocycles such that
\begin{equation}
\label{SGFCocModifFor}
\m_1\left((\vp,\a),\,(\psi,\b)\right)=
\int_{S^1\times{}S^1}2\,\vp\,\psi_{xx}\,dxdy
\end{equation}
(cf. (\ref{MCocy})) and
\begin{equation}
\label{SCMCocModifFor}
\m_2\left((\vp,\a),\,(\psi,\b)\right)=
\int_{S^1\times{}S^1}
\left(
\vp\,\b_{y}+\psi\,\a_y
\right)\,dxdy.
\end{equation}
This defines a two-dimensional central extension $\widehat{\fG}$.

Let us calculate the coadjoint action of $\widehat{\fG}$.
First, observe, that the (regular) dual space $\widehat{\fG}^*$ is isomorphic to
$\widehat{\fG}$. The corresponding Souriau-type cocycles extending (\ref{SouEFor}) to
$\widetilde{\fG}$ can also be easily calculated:
\begin{equation}
\label{SSouEFor}
\begin{array}{rcl}
\displaystyle
S_1(\vp,\a)&=&
\displaystyle
2\,\vp_{xx}\,dx^{\frac{3}{2}},\\[12pt]
\displaystyle
S_2(\vp,\a)&=&
\displaystyle
-\vp_y\,dx^{-\frac{1}{2}}-\a_y\,dx^{\frac{3}{2}},
\end{array}
\end{equation}
so that one can write down the explicit formula of the coadjoint action $\widehat{\ad}^*$ of
$\widehat{\fG}$.
Finally, let us consider only the restriction of $\widehat{\ad}^*$ to the odd part
$\widetilde{\fg}_1$ and apply it to the even part of $\widetilde{\fG}^*$ which is of course
isomorphic to $\widetilde{\fg}$. More precisely, one obtains the map
$$
\widehat{\ad}^*:
\widetilde{\fg}_1\to\End(\widetilde{\fg})
$$
One obtains the following

\begin{prop}
\label{SKPop}
One has 
$$
\widehat{\ad}^*_{
\left(
\begin{array}{c}
\vp\\
\a
\end{array}
\right)
}\left(g,\,v\right)
=
A\left(
\begin{array}{c}
\vp\\
\a
\end{array}
\right),
$$
where $A$ is the operator (\ref{MatSLEq}).
\end{prop}

\noindent
This statement clarifies the origin of the linear differential operators
(\ref{MatSLEq}). Proposition \ref{SKPop} also implies Theorem \ref{KSGenThm}.

\medskip

The ``kinematic'' part of our work is complete, we now turn to the ``dynamics''.

\section{Euler equations on $\widehat{\fg}^*$}

In this section we calculate Euler equations associated with the
Lie algebra $\widehat{\fg}$. 
We are particularly interested in the Euler equations which are bi-Hamiltonian.

\subsection{Hamiltonian formalism on $\widehat{\fg}^*$}

Let us recall very briefly the explicit expression of the Hamiltonian vector fields on the
dual space of a Lie algebra in the infinite-dimensional functional case 
(see, e.g., \cite{Dis,GR} for more details).
Given a functional $H$ on $\widehat{\fg}^*$ which is a (pseudo)differential polynomial:
$$
H(g,\,v)=
\int_{S^1\times{}S^1}
h\left(
g,\,v,
\,g_x,\,v_x,\,g_y,\,v_y,
\,\partial^{-1}_xg,\,\partial^{-1}_xv,
\,\partial^{-1}_yg,\,\partial^{-1}_yv,\,g_{xy},\,v_{xy},\ldots
\right)
\,dxdy,
$$
where $h$ is a polynomial in an infinite set of variables.
The Hamiltonian vector field (\ref{EulEq}) with the Hamiltonian $H$ reads
\begin{equation}
\label{GETypeEq}
(g,\,v)_t=
-\widehat{\ad}^*_{\left(
\frac{\d{}H}{\d{}v},\,\frac{\d{}H}{\d{}g}
\right)}(g,\,v),
\end{equation}
where $\frac{\d{}H}{\d{}v}$ and $\frac{\d{}H}{\d{}g}$ are the standard variational
derivatives given by the (generalized) Euler-Lagrange equations. For instance,
$$
\begin{array}{rcl}
\displaystyle
\frac{\d{}H}{\d{}v} &=&
\displaystyle
h_v-
\frac{\partial}{\partial{}x}
\left(
h_{v_x}
\right)-
\frac{\partial}{\partial{}y}
\left(
h_{v_y}
\right)-
\partial^{-1}_x
\left(
h_{\partial^{-1}_xv}
\right)
-
\partial^{-1}_y
\left(
h_{\partial^{-1}_yv}
\right)\\[14pt]
&&
\displaystyle
+\frac{\partial^2}{\partial{}x^2}
\left(
h_{v_{xx}}
\right)
+\frac{\partial^2}{\partial{}x\partial{}y}
\left(
h_{v_{xy}}
\right)
+\frac{\partial^2}{\partial{}y^2}
\left(
h_{v_{yy}}
\right)
\pm\cdots
\end{array}
$$
where, as usual, $h_v$ means the partial derivative $\frac{\partial{}h}{\partial{}v}$,
similarly $h_{v_x}=\frac{\partial{}h}{\partial{}v_x}$.

\subsection{An Euler equation on 
$\widehat{\fg}^*$ generalizing KP}

Our first example is very close to the classic KP equation (\ref{KP}).
Consider the following quadratic Hamiltonian on
$\widetilde{\fg}^*$:
\begin{equation}
\label{FirstKPHam}
H(g,\,v)=
\int_{S^1\times{}S^1}
\Big(
\frac{v^2}{2}+v\,\partial^{-1}_xg_y
\Big)\,dxdy.
\end{equation}

The variational derivatives of $H$ can be easily computed
$$
\begin{array}{rcl}
\displaystyle
\frac{\d{}H}{\d{}v}(g,\,v)&=&
\displaystyle
\left(
v+\partial^{-1}_xg_y
\right)
\frac{\partial}{\partial x},
\\[12pt]
\displaystyle
\frac{\d{}H}{\d{}g}(g,\,v)&=&
\displaystyle
\partial^{-1}_x
\left(
v_y
\right)
dx^2.
\end{array}
$$
We then use formula (\ref{GETypeEq}) and apply formula (\ref{CExtnEFor}) of the coadjoint
action to obtain the following result.

\begin{prop}
\label{MainThm}
The Euler equation associated with the Lie algebra $\widehat{\fg}$ and the Hamiltonian $H$ is
the following system
\begin{equation}
\label{KPTypeEq}
\begin{array}{l}
\displaystyle
g_t =
vg_x-v_xg-g_yg+g_x\,\partial^{-1}_xg_y
+c_2\,v_y+c_2\,\partial^{-1}_xg_{yy}
\\[10pt]
\displaystyle
v_t =
3v_xv+
c_1v_{xxx}
+c_2\,\partial^{-1}_xv_{yy}
+2vg_y-v_yg
+v_x\,\partial^{-1}_xg_y-2g_x\,\partial^{-1}_xv_y
+c_1g_{xxy}.
\end{array}
\end{equation}
with two indeterminates,  $g(t,x,y)$ and $v(t,x,y)$.
\end{prop}

Note that the second equation in (\ref{KPTypeEq}) can be written as
$$
v_t=3v_xv+c_1v_{xxx}+c_2\,\partial^{-1}_xv_{yy}
+(\hbox{linear terms in $g$}),
$$
and this is nothing but the KP equation with some extra terms in $g$.
In this sense, one can speak of the system (\ref{KPTypeEq}) as a ``generalized KP equation''.

\begin{rmk}
{\rm
If one sets $g\equiv0$ in (\ref{KPTypeEq}), then the first equation gives $v_y=0$. The
function $v$ is thus independent of $y$ and the second equation coincides with KdV (trivially
``looped'' by an extra variable $y$ which does not intervene in the derivatives).
}
\end{rmk}

Unfortunately, we do not know whether the equation (\ref{KPTypeEq}) is
bi-Hamiltonian and have no information regarding its integrability.

\subsection{``Bi-Hamiltonian formalism'' on the dual of a Lie
algebra}

The notion of \textit{integrability} of Hamiltonian systems on infinite-dimensional
(functional) spaces can be understood in a number of different ways.
A quite popular way to define integrability is related to the notion of bi-Hamiltonian
systems that goes back to F. Magri \cite{Mag}.
The best known infinite-dimensional example is the KdV equation.

Let us recall the standard way to obtain bi-Hamiltonian vector fields on the dual of a Lie
algebra. 
Given a Lie algebra $\fa$, the canonical \textit{linear} Poisson structure on $\fa^*$ is
given by
$$
\left\{
F,G
\right\}(m)=
\langle
\left[
d_mF,d_mG
\right],m
\rangle,
$$
where the differentials $d_mF$ and $d_mG$ at a point $m\in\fa^*$ are understood as elements of
the Lie algebra $\fa\cong(\fa^*)^*$.
Consider a \textit{constant} Poisson structure: fix a point
$m_0$ of the dual space and set
$$
\left\{
F,G
\right\}_0(m)=
\langle
\left[
d_mF,d_mG
\right],m_0
\rangle.
$$
It is easy to check that the above Poisson structures are \textit{compatible} (or form a
Poisson pair), i.e., their linear combination
\begin{equation}
\label{Compat}
\{\,,\,\}_{\l}=\{\,,\,\}_0-\l\,\{\,,\,\}
\end{equation}
is again a Poisson structure for all $\l\in\bbC$.

A function $F$ on $\fa^*$ defines now two Hamiltonian vector fields associated with $F$:
$$
m_t=\ad^*_{d_mF}\,m
\qquad
\hbox{and}
\qquad
m_t=\ad^*_{d_mF}\,m_0
$$
corresponding to the first and the second Poisson structure,
respectively. 

\begin{defi}
{\rm
A vector field $X$ on $\fa^*$ is called bi-Hamiltonian if there are two functions, $H$ and
$H'$ such that $X$ is a Hamiltonian vector field of $H$ with respect to
the Poisson structure $\{\,,\,\}$ and is a Hamiltonian vector field of
$H'$ with respect to $\{\,,\,\}_0$.
}
\end{defi}

Bi-Hamiltonian vector fields provides a rich source of integrable systems.
Let $H$ be a function on $\fa^*$ which is a \textit{Casimir function} of the Poisson structure
(\ref{Compat}). This means, for every function $F$, one has
\begin{equation}
\label{Casim}
\{H,\,F\}_{\l}=0
\end{equation}
Assume that $H$ is written in a form of a series
\begin{equation}
\label{Her}
H=H_0+\l\,H_1+\l^2\,H_2+\cdots
\end{equation}
One immediately obtains the following

\begin{prop}
\label{GenCasPro}
The condition (\ref{Casim}) is equivalent to the following two conditions:

(i) $H_0$ is a Casimir function of $\{\,,\,\}_0$.

(ii) For all $k$, the Hamiltonian vector field of $H_{k+1}$ with respect to $\{\,,\,\}_0$
coincides with the Hamiltonian vector field of $H_{k}$ with respect to $\{\,,\,\}$.
\end{prop}

Furthermore, all the Hamiltonians $H_k$ are in involution with respect to both Poisson
structures:
$$
\{H_k,\,H_\ell\}=\{H_k,\,H_\ell\}_0=0
$$
and the corresponding Hamiltonian vector fields commute with each other.
Indeed, if $k\geq\ell$, then one has
$\{H_k,\,H_\ell\}_0=\{H_{k-1},\,H_\ell\}=\{H_{k-1},\,H_{\ell+1}\}_0$ untill one obtains an
expression of the form $\{H_s,\,H_s\}$ or $\{H_s,\,H_s\}_0$ which is identically zero. 

In practice, to construct an integrable hierarchy, one chooses a Casimir function $H_0$ of
the Poisson structure $\{\,,\,\}_0$ and then considers its Hamiltonian vector field with
respect to $\{\,,\,\}$. Thanks to the compatibility condition (\ref{Compat}), this field is
Hamiltonian also with respect to the Poisson structure $\{\,,\,\}_0$ with some Hamiltonian
$H_1$. Then one iterates the procedure.

The above method has been successfully applied to the KdV equation viewed as a Hamiltonian
field on the dual of the Virasoro algebra.

\subsection{Bi-Hamiltonian Euler equation on $\widehat{\fg}^*$}

There exists an Euler equation on $\widehat{\fg}^*$ which is
bi-Hamiltonian. This Euler equation is closely related to the BKP equation (\ref{BKP}) in the
special case $\a=0$.

\begin{thm}
\label{BiHamThm}
The following Hamiltonian
\begin{equation}
\label{SecKPHam}
H(g,\,v)=
\int_{S^1\times{}S^1}
g\,
\Big(
\partial^{-1}_xv_y-
\frac{1}{2}\,\partial^{-1}_xg_y+
\frac{1}{2}\,\frac{c_1}{c_2}\,g_{xx}
\Big)\,dxdy,
\end{equation}
defines a bi-Hamiltonian system on $\widehat{\fg}^*$.
\end{thm}

\begin{proof}
Let us fix the following point of $\widehat{\fg}^*$:
$$
\left(
g,\,u,\,c_1,\,c_2
\right)_0=
\Big(-\frac{\partial}{\partial x},\,
dx^2,\,0,\,0\Big)
$$
and show that the Euler vector field (\ref{SecKPHam}) is also Hamiltonian 
with respect to the constant Poisson structure $\{\,,\,\}_0$.
Consider the following Casimir function of the constant Poisson structure
$\{\,,\,\}_0$
$$
H_0=
\int_{S^1\times{}S^1}
\left(v-g\right)\,dxdy.
$$
Its Hamiltonian vector field with respect to the linear structure $\{\,,\,\}$ is
$$
\left\{
\begin{array}{rcl}
\displaystyle
g_t &=&
g_x\\[6pt]
\displaystyle
v_t &=&
2\,g_x+v_x.
\end{array}
\right.
$$
The compatibility condition (\ref{Compat}) guarantees that this vector field
is Hamiltonian with respect to the constant structure $\{\,,\,\}_0$. Its Hamiltonian
function can be easily computed:
$$
H_1=
\int_{S^1\times{}S^1}
v\,g\,dxdy.
$$
Iterating this procedure, consider its Hamiltonian field with respect to the linear
structure $\{\,,\,\}$:
$$
\left\{
\begin{array}{rcl}
\displaystyle
g_t &=&
c_2\,g_y\\[6pt]
\displaystyle
v_t &=&
c_1\,g_{xxx}+c_2\,v_y.
\end{array}
\right.
$$
Its Hamiltonian with respect to the constant structure $\{\,,\,\}_0$ is proportional to the
function (\ref{SecKPHam}), namely
$
H_2=c_2\,H.
$
We proved that the Hamiltonian $H$ belongs to the Hierarchy (\ref{Her}).
\end{proof}

\begin{rmk}
{\rm
Note that the Hamiltonian $H_1$ is nothing but the quadratic form (\ref{QF}).
In the case of the Lie algebra $\widetilde{\fg}$ this is the Casimir function with
identically zero Hamiltonian vector field.
This is why, in the case of $\widehat{\fg}$, the corresponding
Hamiltonian vector field linearly depends on the central charges $c_1,c_2$.
}
\end{rmk}

Let us now calculate the explicit formula of the Euler equation.

\begin{prop}
\label{VNProPE}
The Euler equation with the Hamiltonian (\ref{SecKPHam}) is of the form:
\begin{equation}
\label{NVTypeEq}
\begin{array}{rcl}
\displaystyle
g_t &=&
g_x\,\partial^{-1}_xg_y-g_y\,g
+c_2\,\partial^{-1}_x\,g_{yy},
\\[12pt]
\displaystyle
v_t &=&
2v\,g_y-v_y\,g
+v_x\,\partial^{-1}_xg_y-2g_x\,\partial^{-1}_xv_y
+g_y\,g+2g_x\,\partial^{-1}_xg_y
-\frac{c_1}{c_2}\left(g_{xxx}\,g+2g_{xx}\,g_x\right)
\\[6pt]
&&
+
2c_1\,g_{xxy}
+c_2
\left(
\partial^{-1}_xv_{yy}
+\partial^{-1}_xg_{yy}
\right).
\end{array}
\end{equation}
\end{prop}
\begin{proof}
We compute the variational derivatives of $H$:
$$
\begin{array}{rcl}
\displaystyle
\frac{\d{}H}{\d{}v}(g,\,v)&=&
\displaystyle
\partial^{-1}_x
\left(
g_y
\right)
\frac{\partial}{\partial x},
\\[12pt]
\displaystyle
\frac{\d{}H}{\d{}g}(g,\,v)&=&
\displaystyle
\left(
\partial^{-1}_xv_y
-\partial^{-1}_xg_y+
\frac{c_1}{c_2}\,g_{xx}
\right)
dx^2.
\end{array}
$$
and then use formula (\ref{GETypeEq}) for the Euler equation together with formula
(\ref{CExtnEFor}) for the coadjoint action.
\end{proof}

The first equation in (\ref{NVTypeEq}) is precisely equation (\ref{OREqn})
already defined in the Introduction.
In the complex case, it is equivalent to BKP (\ref{BKP}) with $\a=0$.

It is an amazing fact that the KP equation (in the preciding section) and the BKP equation
naturally appear in our context on mutually ''dual functions'', namely on $v$ and $g$.

Let us now show how the bi-Hamiltonian technique implies the existence of an
infinite hierarchy of commuting flows.

\subsection{Integrability of equation (\ref{OREqn})\label{FinSec}}

A corollary of Theorem \ref{BiHamThm} is the existence of an infinite series of first integrals
in involution for the field (\ref{NVTypeEq}).
It turns out that the corresponding Hamiltonians are of a particular form.

\begin{prop}
\label{LiLem}
Each Hamiltonian $H_k$ of the constructed hierarchy is an affine functional in $v$, that is
$
H_k(g,\,v)=H'_k(g,\,v)+H''_k(g)
$
and $H'_k$ is linear in $v$.
\end{prop}
\begin{proof}
Let us show that the variational derivative $\frac{\d{}H_k}{\d{}v}$ does not depend on
$v$. 
By construction, the expression for the Hamiltonian vector field of $H_k$ with respect to the
Poisson structure $\{\,,\,\}$ gives
\begin{equation}
\label{Thend}
g_t=
\frac{\d{}H_k}{\d{}v}\,g_x-
\Big(\frac{\d{}H_k}{\d{}v}\Big)_x\,g+
c_2\,\Big(\frac{\d{}H_k}{\d{}v}\Big)_y.
\end{equation}
On the other hand, the same vector field is Hamiltonian field of $H_{k+1}$ with respect to the
Poisson structure $\{\,,\,\}_0$. One has
$$
g_t=\left(\frac{\d{}H_{k+1}}{\d{}v}\right)_x
$$
that expresses the variational derivative $\frac{\d{}H_{k+1}}{\d{}v}$ in terms of the function
$\frac{\d{}H_k}{\d{}v}$.
One then proceeds by induction.
\end{proof}

It follows that the equations (\ref{Thend}) never depend on $v$.
The flows of these vector fields commute with each other since the corresponding Hamiltonian
fields on $\widehat{\fg}$ commute.

\begin{exe}
\label{HerEx}
{\rm
The next vector field of our hierarchy is already quite complicated:
$$
\begin{array}{rcl}
g_t &=&
g
\left(
g\,g_y-g_x\,\partial^{-1}_xg_y
\right)
-
g_x\,\partial^{-1}_x\left(
g\,g_y-g_x\,\partial^{-1}_xg_y
\right)
\\[8pt]
\displaystyle
&&-c_2
\left(
g\,\partial^{-1}_x\,g_{yy}
-g_x\,\partial^{-2}_x\,g_{yy}
+\partial^{-1}_x
\left(
g\,g_y-g_x\,\partial^{-1}_xg_y
\right)_y
\right)
\\[8pt]
\displaystyle
&&+c_2^2\,\partial^{-2}_x\,g_{yyy}.
\end{array}
$$
This is the first higher order equation in the hierarchy of (\ref{OREqn}).
}
\end{exe}

We can now give a partial answer to the question of T. Ratiu.
The dispersionless BKP equation can be realized as an Euler equation
on the dual of the looped cotangent Virasoro algebra.
However, the problem remains open in the case of the classic KP equation.

\section*{Appendix}

All the Lie algebras considered in this paper are generalizations of the
Virasoro algebra with two space variables. 
However, these algebras themselves have interesting generalizations. 
These structures seem to be quite rich and deserve a further study.

\subsection*{Natural generalizations of $L\left(\Vect(S^1)\right)$}

Consider the 2-torus $\bbT^2=S^1\times{}S^1$ parameterized with variables $x$ and $y$. 
Let $\Vect(\bbT^2)$ be the Lie algebra of tangent vector fields on $\bbT^2$.

The Lie algebra $L\left(\Vect(S^1)\right)$ is naturally embedded to $\Vect(\bbT^2)$ as the
Lie subalgebra of vector fields tangent to the constant field 
$X=\frac{\partial}{\partial x}$. This fact suggests the following generalization. Let $V$ be a
compact orientable manifold with a fixed volume form $\om$ and $X\in\Vect(V)$ be a
non-vanishing vector field on
$V$ such that $\Div{}X=0$. We will denote by $\cA_X$ the Lie algebra of vector fields
collinear to $X$; the Lie bracket of $\cA_X$ can be written as follows
$$
[f\,X,g\,X]=
\left(
f\,L_X(g)-g\,L_X(f)
\right)X.
$$
This is clearly a Lie subalgebra of $\Vect(V)$ generalizing $L\left(\Vect(S^1)\right)$. Some
particular cases, such as iterated loop were considered in \cite{RSS}. One can easily
construct a generalization of the Gelfand-Fuchs cocycle (\ref{GFCoc}) on $\cA_X$:
$$
c(f\,X,g\,X)=
\int_Vf
\left(
(L_X)^3g
\right)\,\om.
$$
It is easy to check that this cocycle is non-trivial so that $H^2(\cA_X;\bbC)$ is not
trivial. However, we have no further information about this cohomology group. The geometry of
coadjoint orbits of $\cA$, as well as possible applications to dynamical systems, also remains
an interesting open problem.

\begin{rmk}
{\rm
The condition $\Div{}X=0$ is assumed here mainly for technical reasons (it makes the
formul{\ae} nicer); however, one may think of dropping this condition as well as the condition
on
$X$ to be non-vanishing. }
\end{rmk}

\subsection*{A 2-parameter deformation of $\widehat{\fg}$}

Let us describe a 2-parameter family of Lie algebras which can
be obtained as a deformation of $\widehat{\fg}$.
In \cite{OR} we classified the \textit{non-central} extensions of $\Vect(S^1)$ by the space
of quadratic differentials. The result is as follows.

There are exactly two (up to isomorphism) non-trivial extensions of $\Vect(S^1)$ by $\cF_2$
defined by the following 2-cocycles
$$
\label{CocRhO}
\begin{array}{rcl}
\displaystyle
\rho_1\Big(
f\,\frac{\partial}{\partial x},\,g\,\frac{\partial}{\partial x}
\Big)
&=&
\displaystyle
\left(
f_{xxx}\,g-f\,g_{xxx}
\right)dx^2,\\[12pt]
\displaystyle
\rho_2\Big(
f\,\frac{\partial}{\partial x},\,g\,\frac{\partial}{\partial x}
\Big)
&=&
\displaystyle
\left(
f_{xxx}\,g_x-f_x\,g_{xxx}
\right)dx^2
\end{array}
$$
from $\Vect(S^1)$ to $\cF_2$.

The 2-cocycles $\rho_1,\rho_2$ give rise to the following modification of the Lie algebra law
(\ref{CommutHatEq}).
We set
$$
\begin{array}{rcl}
\displaystyle
\left[
f\,\frac{\partial}{\partial x}+u\,dx^2,\,
g\,\frac{\partial}{\partial x}+v\,dx^2
\right]_{(\kappa_1,\kappa_2)}
&=&
\displaystyle
\left[f\,\frac{\partial}{\partial x}+u\,dx^2,\,
g\,\frac{\partial}{\partial x}+v\,dx^2\right]\\[12pt]
\displaystyle
&+&
\displaystyle
\kappa_1\,\rho_1\Big(
f\,\frac{\partial}{\partial x},\,g\,\frac{\partial}{\partial x}
\Big)
+\kappa_2\,\rho_2\Big(
f\,\frac{\partial}{\partial x},\,g\,\frac{\partial}{\partial x}
\Big),
\end{array}
$$
where $\kappa_1\,\kappa_2\in\bbC$ are parameters.

This deformed commutator satisfies the Jacobi identity and provides an interesting Lie
algebra structure.

\vskip 0.5cm

\textbf{Acknowledgments}.
We are grateful to T. Ratiu for his interest in this work.
We also wish to thank B. Khesin and A. Reiman for enlightening discussions.

\vskip 0.5cm


Institut Camille Jordan

Universit\'e Claude Bernard Lyon 1,

21 Avenue Claude Bernard,

69622 Villeurbanne Cedex,

FRANCE;

ovsienko@math.univ-lyon1.fr,

roger@math.univ-lyon1.fr

\end{document}